\documentclass[aps,pra,twocolumn,superscriptaddress,showpacs]{revtex4}
\usepackage{amssymb}
\usepackage{amsmath}
\usepackage{graphics}
\usepackage{color}

\definecolor{red}{rgb}{1,0,0}
\input{tcilatex}

\begin{document}

\title{Influence of the polarization in grazing scattering of fast helium
atoms from LiF(001) surfaces.}
\author{M.S. Gravielle}
\author{J.E. Miraglia}
\affiliation{Instituto de Astronom\'{\i}a y F\'{\i}sica del Espacio, CONICET, Casilla de
Correo 67, Sucursal 28, 1428 Buenos Aires, Argentina, and\\
Dpto. de F\'{\i}sica, FCEN, Universidad de Buenos Aires, Buenos Aires,
Argentina.}
\date{\today }

\begin{abstract}
Grazing scattering of neutral atoms from insulator surfaces is investigated
in the intermediate velocity range, in which interference effects have been
recently observed. To describe this process we introduce a distorted-wave
method, based on the use of the eikonal wave function, which takes into
account the phase of the scattering state along the classical projectile
path. The eikonal theory is applied to evaluate the angular distribution of
few keV helium atoms after impinging on a LiF(001) surface along low-index
crystallographic directions. The interest focuses on the role played by the
projectile polarization produced by cations and anions of the crystal
surface. For the considered collision system we found a polarization
channel, corresponding to the direction \TEXTsymbol{<}110\TEXTsymbol{>},
which is affected by this effect, while for incidence in the direction 
\TEXTsymbol{<}100\TEXTsymbol{>} the polarization contribution is nearly
negligible. The proposed eikonal approach, including polarization effects,
provides angular projectile spectra in good agreement with the experimental
data.
\end{abstract}

\maketitle

\section{Introduction}

Experimental evidences of \ interference effects produced during grazing
scattering of fast atoms from insulator surfaces were recently presented in
Refs. \cite{Winter07,Roncin07,Winter08}. Under axial surface channeling
conditions, reported measurements of scattered projectile distributions
display well defined spots, associated with diffraction patterns originated
by the periodic structure of the crystal. Even though the diffraction of
particles from crystal surfaces has been a well-understood phenomena from
the beginning of quantum mechanics, the importance of these experimental
results is due to the effect not being expected to be observable for light
atoms with energies in the keV range, whose de Broglie wavelengths are some
orders of magnitude smaller than the shortest interatomic distance in the
crystal.

Two different interference mechanisms have been proposed to explain the
experimental observations. The first of them \cite{Winter07,Roncin07} is
related to the diffraction by a periodic lattice and relies on the
assumption that the projectile motion perpendicular to the axial incidence
channel can be decoupled from the parallel one. The semiclassical approach 
\cite{Roncin07} proposed for the description of this mechanism predicts a
maximum in the distribution when the component normal to the scattering
plane of the final projectile momentum coincides with a reciprocal lattice
vector. Such a result, also confirmed by means of a wave packet propagation
approach \cite{Roncin07}, is in accord with the experimental data for the
lowest impact energies ~\cite{Winter07,Roncin07}. The second mechanism \cite%
{Winter08}, called supernumerary rainbow, is originated by the corrugation
of the surface potential, which gives rise to a quantum interference between
projectiles emerging from the surface with the same direction but reflected
at different turning points ~\cite{Avrin94}.

To describe the experimentally observed patterns, in this article we
introduce a distorted-wave model, named surface eikonal approximation, which
is valid for small de Broglie wavelengths of incident atoms. This method
makes use of the eikonal wave function \cite{Joachain} to represent the
elastic collision with the surface, while the projectile movement is
classically described, taking into account different initial conditions. The
surface eikonal approach can be considered as an extension of the well-known
Glauber approximation \cite{Glauber59} for collisions with corrugated
surfaces \cite{Manson00} instead of atoms, but considering axial channeled
trajectories. It includes both interference mechanisms and the idea behind
it essentially coincides with that of the semiclassical formalism \cite%
{Avrin94} used in Ref. \cite{Winter08}.

The surface eikonal approach is here employed to describe angular
distributions of \ swift\ He$^{0}$ atoms scattered off from a LiF(001)
surface, for which there are experimental data available \cite%
{Winter07,Roncin07,Winter08}. As the considered process is very sensitive to
the description of the surface potential, the aim of the work is to
investigate the influence of the polarization on the interference patterns.
In our model the interaction of the incident atom with the crystal surface
is represented as a sum of individual interatomic potentials, which take
into account the contribution of \ the different ionic centres of the
insulator material \cite{Garcia07}. To evaluate the interatomic potentials
we use the Abrahamson approximation \cite{Abrahamson}, adding the asymptotic
contribution of the projectile polarization. The role of the polarization is
analyzed for incidence along the \TEXTsymbol{<}100\TEXTsymbol{>} and 
\TEXTsymbol{<}110\TEXTsymbol{>} channels, finding that polarization effects
are important for this latter crystallographic direction. Atomic units ($%
e^{2}=\hbar =m_{e}=1$) are used unless otherwise stated.\medskip

\section{Surface Eikonal approximation}

Let us consider the grazing impact of an atomic projectile ($P$), with mass $%
m_{P}$, on a crystal surface ($S$). As a result of the collision, the
projectile with initial momentum $\vec{K}_{i}$ is elastically scattered from
the surface, ending in a final state with momentum $\vec{K}_{f}$. The frame
of reference is fixed on a target ion belonging to the first atomic layer,
with the surface contained in the $x-y$ plane and the $\hat{z}$ versor
perpendicular to the surface, aiming towards the vacuum region.

We assume that the state $\Psi _{i}^{+}$ associated with the collision
system satisfies the time-independent Schr\"{o}dinger equation for the
Hamiltonian 
\begin{equation}
H=-\frac{1}{2m_{P}}\nabla _{\vec{R}_{P}}^{2}+V_{SP}(\vec{R}_{P}),  \label{H}
\end{equation}%
where $\vec{R}_{P}\ $denotes the position of the center of mass of the
incident atom and $V_{SP}$ is the surface-projectile interaction. As initial
condition, when the projectile is far from the surface, $\Psi _{i}^{+}$
tends to the state $\Phi _{i}$, with $\Phi _{j}(\vec{R}_{P})=(2\pi
)^{-3/2}\exp (i\vec{K}_{j}\cdot \vec{R}_{P})$, $j=i(f)$, the initial (final)
unperturbed wave function.

The central magnitude to describe the elastic scattering process is the
transition matrix, which reads 
\begin{equation}
T_{if}=\int d\vec{R}_{P}\ \Phi _{f}^{^{\ast }}(\vec{R}_{P})\ V_{SP}(\vec{R}%
_{P})\Psi _{i}^{+}(\vec{R}_{P}).  \label{Tif}
\end{equation}%
In the energy range of interest, Eq. (\ref{Tif}) can be expressed in terms
of the classical trajectory of the projectile - $\mathcal{\vec{R}}_{P}$ - \
by means of the substitution $\vec{R}_{P}\cong \mathcal{\vec{R}}_{P}$, like
in the usual semiclassical formalism \cite{McDowellColeman}. The position $%
\mathcal{\vec{R}}_{P}$ of the incident atom at a given time $t$ is governed
by the Newton equations associated with the potential $V_{SP}$, verifying
the relation $\mathcal{\vec{R}}_{P}(\vec{R}_{os},t)=\vec{R}_{os}+Z_{o}\hat{z}%
+\int\nolimits_{-\infty }^{t}dt^{\prime }$ $\vec{v}(\vec{R}_{os},t^{\prime
}) $, where $\vec{v}(\vec{R}_{os},t)$ is the classical velocity of the
projectile, $\vec{R}_{os}=(X_{o},Y_{o},0)$ identifies its initial position
on the surface plane and $Z_{o}\rightarrow +\infty $. A sketch picture of
the projectile path and the coordinate system is displayed in Fig. 1. By
replacing the integration variables $\vec{R}_{P}=(X_{P},Y_{P},Z_{P})$ by the
new ones $\{X_{o},Y_{o},t\}$ in Eq. (\ref{Tif}), the transition matrix is
expressed as \cite{Hubbard83}

\begin{eqnarray}
T_{if} &=&\int d\vec{R}_{os}\ \int\limits_{-\infty }^{+\infty }dt\ \
\left\vert v_{z}(\vec{R}_{os},t)\right\vert \times  \notag \\
&&\Phi _{f}^{^{\ast }}(\mathcal{\vec{R}}_{P})\ V_{SP}(\mathcal{\vec{R}}%
_{P})\Psi _{i}^{+}(\mathcal{\vec{R}}_{P}),  \label{Tif-ipn}
\end{eqnarray}
where $v_{z}(\vec{R}_{os},t)$ is the component of $\vec{v}$ normal to the
surface.

Since the de Broglie wavelength of the incident projectile, $\lambda =2\pi
/K_{i}$, is sufficiently short compared with the characteristic distance of
the surface potential, we approximate the scattering state $\Psi _{i}^{+}$
by means of the eikonal wave function \cite{Joachain}, i.e. 
\begin{equation}
\Psi _{i}^{+}(\mathcal{\vec{R}}_{P})\simeq \chi _{i}^{^{(eik)+}}(\mathcal{%
\vec{R}}_{P})=\Phi _{i}(\mathcal{\vec{R}}_{P})\exp (-i\eta (\mathcal{\vec{R}}%
_{P})),  \label{feik}
\end{equation}%
where $\eta (\mathcal{\vec{R}}_{P})$ is the eikonal phase, defined as 
\begin{equation}
\eta (\mathcal{\vec{R}}_{P}(\vec{R}_{os},t))=\int\limits_{-\infty
}^{t}dt^{\prime }\ V_{SP}(\mathcal{\vec{R}}_{P}(\vec{R}_{os},t^{\prime })).
\label{fasen}
\end{equation}%
By introducing the function $\chi _{i}^{^{(eik)+}}$ in Eq. (\ref{Tif-ipn})
the eikonal transition matrix reads 
\begin{eqnarray}
T_{if}^{(eik)} &=&\frac{1}{(2\pi )^{3}}\int d\vec{R}_{os}\
\int\limits_{-\infty }^{+\infty }dt\ \left\vert v_{z}(\vec{R}%
_{os},t)\right\vert \times  \notag \\
&&\exp [-i\vec{Q}\cdot .\mathcal{\vec{R}}_{P}-i\eta (\mathcal{\vec{R}}%
_{P})]\ V_{SP}(\mathcal{\vec{R}}_{P}),  \label{Teikn}
\end{eqnarray}%
where $\vec{Q}=\vec{K}_{f}-\vec{K}_{i}$ is the projectile momentum transfer
and the final momentum $\vec{K}_{f}$ satisfies the energy conservation, i.e. 
$K_{f}=K_{i}$. The differential probability, per unit of surface area, of
elastic scattering with final momentum $\vec{K}_{f}$ in the direction of the
solid angle $\Omega _{f}$ is obtained from Eq. (\ref{Teikn}) as $dP/d\Omega
_{f}=(2\pi )^{4}m_{P}^{2}\left\vert \tilde{T}_{if}^{(eik)}\right\vert ^{2}$,
where $\tilde{T}_{if}^{(eik)}$ denotes the eikonal T-matrix element,
normalized per unit area. Note that the main difference between the usual
eikonal scattering amplitude \cite{Joachain, Manson00} and Eq. (\ref{Teikn})
arises from the use of axial channeled trajectories instead of straight-line
ones.

The first Born T-matrix element can be derived from Eq. (\ref{Teikn}) by
neglecting the eikonal phase; that is, by fixing $\eta (\mathcal{\vec{R}}%
_{P})=0$. It reads 
\begin{eqnarray}
T_{if}^{(Born)} &=&\frac{1}{(2\pi )^{3}}\int d\vec{R}_{os}\
\int\limits_{-\infty }^{+\infty }dt\ \left\vert v_{z}(\vec{R}%
_{os},t)\right\vert \times   \notag \\
&&\exp (-i\vec{Q}\cdot \mathcal{\vec{R}}_{P})\ V_{SP}(\mathcal{\vec{R}}_{P}).
\label{Tborn}
\end{eqnarray}

\section{Interaction potentials}

In this work, the projectile-surface potential $V_{SP}$ contains the static
and polarization interactions, i.e., $V_{SP}=V_{SP}^{(st)}+V_{SP}^{(pol)}$.
Due to the insulator character of the surface, both term can be derived by
considering\ the surface as composed by independent target ions.
Consequently, the static potential $V_{SP}^{(st)}$ is expressed as a sum of
individual interatomic potentials, $V_{st}(\vec{R})$, which represent the
static interaction of the incident atom with solid ions placed at different
lattice sites \cite{Garcia07}. Following the Lenz energy functional \cite%
{Lenz,Jensen}, for the two types of target ions - alkali-metal and halide-
the static ion-atom interaction is found to be the sum of three terms,

\begin{equation}
V_{st}(\vec{R})=V_{Coul}(\vec{R})+V_{kin}(\vec{R})+V_{xch}(\vec{R}).
\label{Vat}
\end{equation}

The first term is the well-known electrostatic Coulomb interaction,

\begin{equation}
V_{Coul}(\vec{R})=\frac{1}{2}\int \int d\vec{r}\mathbf{\;}d\vec{r}^{\prime
}D_{T}(\vec{r})\frac{1}{|\vec{r}-\vec{r}^{\prime }|}D_{P}(\vec{r}^{\prime }-%
\vec{R}),  \label{Vcou}
\end{equation}%
where $D_{T}(\vec{r})=Z_{T}\delta (\vec{r})-$ $\mathbf{\rho }_{T}(\vec{r})\;$%
and $D_{P}(\vec{r})=Z_{P}\delta (\vec{r})-$ $\mathbf{\rho }_{P}(\vec{r})\;$%
are the target and projectile charge densities, $\delta $ is the Dirac delta
function, situated at the position of the nucleus, $Z_{T}\;(Z_{P})$ is the
target (projectile) nucleus charge,\ and $\mathbf{\rho }_{T}\;(\mathbf{\rho }%
_{P})$ is the target (projectile) electronic density. Note that $V_{Coul}$
is composed by four terms, including the internuclear and electron-electron
repulsions as well as the attractive electron-nucleus potentials.

By employing the Abrahamson approximation \cite{Abrahamson} the second term
of Eq.(\ref{Vat}), named the kinetic potential, reads

\begin{eqnarray}
V_{kin}(\vec{R})\frac{1}{2.871} &=&\int d\vec{r}\mathbf{\;}\left[ \mathbf{%
\rho }_{T}(\vec{r})+\mathbf{\rho }_{P}(\vec{r}\mathbf{-}\vec{R})\right]
^{5/3}  \label{Vkin} \\
&&-\int d\vec{r}\ \mathbf{\rho }_{T}^{5/3}(\vec{r})-\int d\vec{r}\ \mathbf{%
\;\rho }_{P}^{5/3}(\mathbf{r-}\vec{R}).  \notag
\end{eqnarray}%
This potential is essentially positive and represents the reaction to the
compression of the electronic density, considered as a free-electron gas.
The third term describes the exchange potential within the local
approximation and it reads

\begin{eqnarray}
V_{xch}(\vec{R})\frac{(-1)}{0.738} &=&\int d\vec{r}\mathbf{\;}\left[ \mathbf{%
\rho }_{T}(\vec{r})+\mathbf{\rho }_{P}(\vec{r}\mathbf{-}\vec{R})\right]
^{4/3}  \label{Vex} \\
&&-\int d\vec{r}\mathbf{\;\rho }_{T}^{4/3}(\vec{r})-\int d\vec{r}\mathbf{%
\;\rho }_{P}^{4/3}(\vec{r}\mathbf{-}\vec{R}).  \notag
\end{eqnarray}%
Finally, the potential $V_{SP}^{(pol)}$ takes into account the polarization
of the neutral projectile in the presence of target ions, which is not
included in the original Abrahamson model. Following the usual derivation of
the atomic polarization potential \cite{Joachain}, although in this case,
not for only one perturbative charge but a collection of them, which
represent the different target ions, the asymptotic polarization potential
reads

\begin{equation}
V_{SP}^{(pol)}(\vec{R})=-\frac{\alpha }{2}\sum\limits_{i,j}\frac{%
Z_{Ti}^{(\infty )}}{(R_{0i}^{2}+R_{i}^{2})}(\hat{R}_{i}\cdot \hat{R}_{j})%
\frac{Z_{Tj}^{(\infty )}}{(R_{0j}^{2}+R_{j}^{2})},  \label{Vpol}
\end{equation}%
where the sum formally includes all the \ target ions of the crystal, $%
\alpha $ is the polarizability of the projectile, with $\alpha =1.38$ a.u.
for Helium \cite{Bederson}, and $\vec{R}_{i}$ ($\vec{R}_{j}$) represents the
position vector of the projectile with respect to the target ion labelled as 
$i$ ($j$), with $\hat{R}_{i}=\vec{R}_{i}/R_{i}$. In Eq.(\ref{Vpol}), $%
Z_{Ti}^{(\infty )}$ is the residual charge of the target ion at long
distances, being $Z_{Ti}^{(\infty )}=1$ for Li$^{+}$ and $Z_{Ti}^{(\infty
)}=-1$ for F$^{-}$. At short distances the polarization contribution of the
target ion $i$ is reduced with a cutoff, which is always of the order of the
radius of the atom, that is, $R_{0i}=\left\langle r\right\rangle
_{Ti}+\left\langle r\right\rangle _{P}$, where \ $\left\langle
r\right\rangle _{Ti}$ $\,$($\left\langle r\right\rangle _{P}$) is the target
(projectile) mean radius. We employed the values 1.09, 0.67, and 1.41 a.u.
for the He, Li$^{+}$ and F$^{-}$ mean radii, respectively. Far from the
surface, the diagonal terms ($i=j$) of the polarization potential given by
Eq. (\ref{Vpol}) satisfy the well known behavior $-\alpha /2R^{4}$, but the
extra-diagonal terms ($i\neq j$) are weighted by a directional factor that
depends on the crystal ordering \cite{nota,Celli85}. Notice that as we are
dealing with neutral projectiles, we have not taken into account the dynamic
polarization of the surface ions \cite{Garcia07} because this effect
represents a higher-order correction of the interatomic potential ($%
\varpropto R^{-6}$).

\section{\protect\bigskip Results}

We applied the model to neutral helium atoms impinging grazingly on a LiF
crystal surface under axial surface channeling conditions. The impact energy
ranged from 0.2 to 8.6 keV, \ corresponding to the experiments of Refs. \cite%
{Winter07,Roncin07,Winter08}. In the crystal surface, ions belonging to the
topmost atomic layer were slightly displaced from their equilibrium
positions, in accord with Ref. \cite{Vogt02}.

To describe the projectile-surface potential we employed the punctual model
of Ref. \cite{Garcia07}, evaluating the He-Li$^{+}$ and \ He-F$^{-}$
interatomic potentials from Eq.(\ref{Vat}). Hartree-Fock Slater wave
functions from Clementi-Roetti \cite{ClementiRoetti} were used to calculate
the electronic densities $\mathbf{\;\rho }_{T}$ and $\mathbf{\;\rho }_{P}$.
It allowed us to derive a closed form for $V_{Coul}$, \ while $V_{kin}$ and $%
V_{xch}$ were obtained from numerical integrations. In Fig. 2 it seemed
convenient to plot the scaled expression $W(R)=V(R)\ast R(1+2R^{3})$ for Li$%
^{+}$ and F$^{-}$, respectively, where $V(R)$ includes the static potential
[Eq.(\ref{Vat})] plus the diagonal polarization contribution, i.e. the $i=j$
term of Eq.(\ref{Vpol}). From the figure we can differentiate two different
regions of the interatomic potentials. As $R\rightarrow 0$, $W(R)\rightarrow
Z_{T}Z_{P}$ and the sharp increase at the origin corresponds to the
electrostatic contribution $V_{Coul}(R)$, while the maximum at intermediate
distances\ is mainly due to the statistical contribution, i.e. $%
V_{kin}(R)+V_{xch}(R)$. $\ $\ Note that present static potentials are almost
indistinguishable from the ones of Gordon and Kim \cite{Gordon} (empty
circles), employed in Ref. [3]. The asymptotic limit of $V$ as $R\rightarrow
\infty $ is affected by the polarization, i.e. $V\ast R(1+2R^{3})$ $%
\rightarrow -\alpha .$

The projectile trajectory was derived \ from classical dynamics with the
Runge-Kutta method. At every step we took into account the 4$^{th}$ order
nearest neighbor target ions (i.e. $8\times 8\times 4$), which includes the
interaction of the projectile with the topmost atomic layer and three more
layers below it. We have made sure our results do no depend on the
considered number of nearest neighbors by increasing this number to include
up to 8$^{th}$ order nearest neighbors (i.e., 8 atomic planes).

The evaluation of the eikonal transition matrix involves an integration on
the starting point $\vec{R}_{os}$ of the classical trajectory, which was
calculated with the MonteCarlo technique, varying $\vec{R}_{os}$ on the area
of the unit cell as a consequence of the surface invariance. In every case
we considered around $10^{5}$ classical trajectories with random initial
positions, and this number was varied in order to test the convergency of \
our calculations. The further integration on $t$ involved in Eq. (\ref{Teikn}%
) was numerically solved with a relative error lower than 0.1\%. \ To obtain
the differential probability $dP/d\Omega _{f}$, we have to add the T-matrix
elements corresponding to different values of $\vec{R}_{os}$ that lead to
the \textit{same} final momentum $\vec{K}_{f}$. For this purpose we employed
a grid for the angles $\theta _{f}$ and $\varphi _{f}$ of $100\times 100$
points, where $\theta _{f}$ and $\varphi _{f}$ are the final polar and
azimuthal angles, respectively, of the final momentum $\vec{K}_{f}$. \ In
all calculations we oriented the $\hat{x}$ versor along the low-index
direction of the crystal surface coinciding with the impact direction;
therefore, the azimuthal angle $\varphi _{f}$ \ is measured with respect to
the incidence direction on the surface plane (see Fig. 1). In full accord
with the experiments of Refs. \cite{Winter07,Roncin07,Winter08} we found
that under axial surface channeling conditions the relation $\theta
_{f}^{2}+\varphi _{f}^{2}\approxeq \theta _{i}^{2}$ is almost strictly
verified by all classically scattered projectiles and consequently, the
angular projectile distribution shows the usual banana shape \cite{Meyer95}.

We start the analysis by considering the experimental case of Fig. 5 of \
Ref. \cite{Winter07}; that is, 3 keV $^{3}$He atoms impinging on a LiF(001)
surface along the crystallographic direction \TEXTsymbol{<}110\TEXTsymbol{>}
with a glancing angle ($\theta _{i}=1.1\deg $). This collision system looks
adequate for the eikonal description because the de Broglie wavelength of
the incident atom ($\lambda =0.0057$ a.u.) is almost three orders of
magnitude smaller than the characteristic interatomic distance. In Fig. 3 \
we plot the differential probability $dP/d\varphi _{f}$, as a function of
the azimuthal angle $\varphi _{f}$, multiplying the results by an arbitrary
factor in order to show the different curves separately. The eikonal
spectrum displays strong interference signatures, presenting pronounced
maxima symmetrically placed with respect to the incidence direction, which
corresponds to $\varphi _{f}=0$. This interference pattern can be directly
compared with the experimental spots of Ref. \cite{Winter07}, which are
displayed with stars, numbering them from the central one. The eikonal
distribution nearly agrees with the experimental one, although \ the eikonal
maxima associated to the peaks $\pm $1, $\pm $2, $\pm $3, and $\pm $4 are
slightly shifted to higher values. Notice that the extreme angles of the
eikonal spectrum are related to the rainbow scattering and the corresponding
maxima display a sharp shape. These peaks are also present in the classical
distribution, defined as the number of projectile trajectories reaching a
given final azimuthal angle $\varphi _{f}$, which is shown in an \textit{%
absolute} scale in Fig. 3. The classical scattering distribution presents
the typical rainbow profile \cite{Schuller04}, with only two maxima around
the extremes of the angular spectrum -the rainbow angles. Then, the absence
of intermediate structures\ in the classical spectrum confirms the concept
that interference effects are a consequence of quantum coherence between
projectiles moving along different paths but ending in the same final state.

With the aim of analyzing the influence of the polarization of \ helium
atoms, in Fig. 3 we also plot eikonal values obtained by neglecting the
polarization potential; that is, by dropping $V_{SP}^{(pol)}$ in the
projectile-surface interaction. We found that for incidence along the
direction \TEXTsymbol{<}110\TEXTsymbol{>}, the angular distribution of
scattered atoms is affected by the projectile polarization. When the
polarization is not included in the calculation, the central maximum becomes
a minimum, modifying the total number of peaks displayed by the eikonal
distribution. In turn, the extreme maxima, associated with the rainbow
angles, are only slightly altered by the polarization. Both angular regions
- central and external- of the eikonal spectrum are associated with
different zones of the interatomic potentials\ that are probed by axial
channeled projectiles. He$^{0}$ atoms that reach azimuthal angles $\varphi
_{f}$ near 0 move\ over the ionic rows that form the channel, farther than 2
a.u. from the surface, interacting with the long-distance contribution of
the surface-projectile potential. As a such contribution is dominated by the
term corresponding to the polarization potential, given by Eq.(\ref{Vpol}),
it explains the influence of this effect on the central zone of the
spectrum. Projectiles that end in the rainbow angular region, instead,
suffer closer collisions with F ionic centers, being affected by the
short-distance behavior of the interatomic potentials, which is determined
by coulombic and statistical contributions.

In Fig. 4 we investigate the elastic scattering along the direction 
\TEXTsymbol{<}100\TEXTsymbol{>} by considering a higher impact energy (8.6
keV). The eikonal differential probability is plotted in Fig. 4 (a) as a
function of the final azimuthal angle, together with experimental spots of
Fig. 2 of Ref. \cite{Winter08}. For this collision system, in addition to
the two rainbow maxima, the eikonal distribution presents four similar
peaks, symmetrically placed around $\varphi _{f}=0$, \ and a very small
central maximum. The number of main maxima of the eikonal profile coincides
with that of the experimental pattern \cite{Winter08}, although the
positions of the peaks are again shifted to higher values in comparison with
the experimental ones.\ In all the cases we found that slight changes in the
interatomic potentials produce substantial modifications in the angular
spectrum of scattered projectiles. Hence, discrepancies between theoretical
and experimental spectra could be associated with very subtle differences in
the projectile-surface potential. In Fig. 4 (b) we compare experimental
intensities \cite{Winter08} with eikonal probabilities, now plotted in
linear scale, as function of the deflection angle $\Theta $, defined as $%
\Theta =\arctan (\varphi _{f}\ /\theta _{f})$. Taking into account that our
theoretical results were obtained by considering fixed positions of the
target ions, \textit{without} including the thermal vibration, and they were 
\textit{not} convoluted with experimental conditions, the eikonal model
reproduces fairly well the main features of the experimental spectrum.

In order to investigate the effect of the polarization in the channel 
\TEXTsymbol{<}100\TEXTsymbol{>}, in Fig 4 (a) we show eikonal values derived
by eliminating the polarization potential. Remarkably, eikonal results with
and without including the projectile polarization agree with each other for
incidence along the \TEXTsymbol{<}100\TEXTsymbol{>} direction, indicating
that polarization effects play a minor role in this channel. It is a
consequence of the ordering of the halide and alkali ions involved in the
axial surface channeling. As observed from Eq. (\ref{Vpol}), when the
projectile moves along the channel far from the surface plane, the factors
of the polarization potential coming from \ F$^{-}$ and Li$^{+}$ have
opposite signs and they compensate their contributions to order $r^{-4}$
when F$^{-}$ and Li$^{+}$ ions are placed in front of each other, as it
happens in the \TEXTsymbol{<}100\TEXTsymbol{>} direction. Furthermore,
within a row model, the \TEXTsymbol{<}100\TEXTsymbol{>} rows \ - formed by
alternate cations an anions - display a neutral charge, which reduces the
polarization of the incident atom. In the \TEXTsymbol{<}110\TEXTsymbol{>}
direction, instead, \ not only are there separated cation and anion rows,
with positive and negative net charges respectively, but also Li$^{+}$ and F$%
^{-}$ions are not in front of each other along the channel, which originates
an effective polarization potential. This is the reason why polarization
effects become evident for incidence along the \TEXTsymbol{<}110\TEXTsymbol{>%
} direction but not in the channel \TEXTsymbol{<}100\TEXTsymbol{>}.

Besides, in Fig. 4 (a) we also show the angular distribution obtained within
the first Born approximation [Eq.(\ref{Tborn})], which is derived from Eq. (%
\ref{Teikn}) by eliminating the eikonal phase. The Born profile displays a
different diffraction pattern, with a broad central maximum, not present in
the experiment, indicating that interference structures of the surface
eikonal model are affected by the phase $\eta $, given in Eq. (\ref{fasen}).
However, note that differences between eikonal and Born distributions vary
with the considered collision system.

Finally, in Fig. 5 we considered the incidence conditions of Ref. \cite%
{Roncin07}, which correspond to a smaller impact energy (0.2 keV). Notice
that this energy is close to the limit of validity of the eikonal model,
which is expected to be adequate for high velocities. For scattering along
the direction \TEXTsymbol{<}110\TEXTsymbol{>}, the eikonal differential
probability is displayed in Fig. 5, as a function of the azimuthal angle,
comparing it with the spots of Fig. \ 1 of \ Ref. \cite{Roncin07}. Also in
this case, the agreement of the eikonal theory with the experiment is
reasonable good. Both profiles - eikonal and experimental - present similar
structures, with a central maximum and two additional peaks, not equally
spaced, to each side. However, the experimental peaks $\pm 1$ are narrower
than the eikonal ones, and small structures around the rainbow angles are
absent in the theory, corresponding to the worst disagreement found in the
present work. Discrepancies between the theory and the experiment can be
again attributed to extremely subtle distinctions in the projectile-surface
potential. Moreover, we should mention that the small rumpling ($d=0.037$
a.u.) of \ the surface ions introduced in our model \cite{Vogt02} affects
the interference pattern.

Again, like in Fig. 3, the central zone of the eikonal spectrum of Fig. 5 is
associated with the long-distance behavior of the surface interaction, which
is governed by the projectile polarization. When $V_{SP}^{(pol)}$ is
dropped, the central maximum of the eikonal distribution completely
disappears, in disagreement with the experimental data.

To investigate in detail the central zone of the eikonal spectrum, in Fig 6
we plot the first ten projectile trajectories, provided by the MonteCarlo
code, that contribute to the distribution at the final azimuthal angle $%
\varphi _{f}\cong 0$. For the collision system of \ Fig. 3 we observe that
all the atoms that end in this angular region move just over F$^{-}$ or Li$%
^{+}$ rows. In this case, turning points corresponding to the z- movement
are almost independent \ of the motion perpendicular to the scattering
plane, being approximately situated 2.8 a.u. (2.2 a.u.) above the topmost
atomic layer for projectiles moving over F$^{-}$ (Li$^{+}$) rows. From Fig.
6 (c), the transversal kinetic energy, defined as $E_{\perp
}^{(kin)}=m_{P}(v_{y}^{2}+v_{z}^{2})/2$, slightly increases just before and
after reaching the collision region, indicating that incident atoms are
affected by an attractive polarization potential. The projectile-surface
potential along classical trajectories, shown in Fig. 6 (d), displays an
oscillatory pattern produced by the interaction with the different ionic
centers of the crystal surface. Consequently, the total transversal energy $%
E_{\perp }=E_{\perp }^{(kin)}+V_{SP}$ presents fluctuations along the
classical projectile path. However, the mean value $\left\langle E_{\perp
}\right\rangle =\left\langle E_{\perp }^{(kin)}\right\rangle +\left\langle
V_{SP}\right\rangle $ \ keeps equal to the initial value $%
E_{iz}=m_{P}v_{iz}^{2}/2$ along the whole trajectory, supporting to some
extend the decoupling of the transversal movement from the parallel one,
proposed in Ref. \cite{Roncin07}.

\section{Summary}

In conclusion, we have developed a surface eikonal approach to deal with
interference patterns produced by impact of swift atoms on insulator
surfaces. The proposed method has been applied to few keV He atoms grazing
impinging on LiF(001) along the \TEXTsymbol{<}110\TEXTsymbol{>} and 
\TEXTsymbol{<}100\TEXTsymbol{>} directions. Projectile spectra derived with
the eikonal approximation display well-defined interference structures,
originated by atoms that follow different paths but end scattered with the
same final momentum. As the projectile distribution \ strongly depends on
the description employed to represent the projectile-surface interaction,
the study focused on the influence of the projectile polarization on the
angular spectrum. We conclude that the polarization potential is essential
to describe the elastic scattering along the \TEXTsymbol{<}110\TEXTsymbol{>}
channel, while in the direction \TEXTsymbol{<}100\TEXTsymbol{>} its
contribution is negligible. Angular spectra derived from the eikonal model,
including the polarization effect, are in concordance with the available
experimental \ data \cite{Winter07,Roncin07,Winter08}. But a better
representation of the surface potential, taking into account that target
ions are part of a surface, might modify the present results. Then, by
including more precise electronic densities this method may be useful to
investigate very delicate items, such as long-distance potentials or crystal
ion displacements, which are difficult to make evident experimentally \cite%
{Farias04}.

\begin{acknowledgments}
This work was supported by CONICET, UBA, and ANPCyT of Argentina.
\end{acknowledgments}

\begin{figure}[tbp]
\suppressfloats
\caption{ Schematic depiction of the coordinate system.}
\end{figure}

\begin{figure}[tbp]
\suppressfloats
\caption{ Scaled interatomic potentials for (a) He-Li$^{+}$ and (b) He-F$^{-}
$. Solid line, potential including the diagonal (\textit{i=j}) polarization
contribution; dashed line, static potential, without polarization; and empty
circles, results reported by Gordon and Kim in Ref. [19]. }
\end{figure}

\begin{figure}[tbp]
\suppressfloats
\caption{Azimuthal angular distribution of elastic scattered projectiles for
3 keV $^{3}$He atoms impinging on LiF(001) along the direction \TEXTsymbol{<}%
110\TEXTsymbol{>}, with the incidence angle $\protect\theta _{i}=1.1\deg $.
Solid line, differential probability derived from the surface eikonal
approach, including polarization effects; dashed line, surface eikonal
results without including the projectile polarization. Full stars,
experimental spots of Fig. 5 of Ref. [1]. Empty circles, classical
distribution, as explained in the text, in \textit{absolute} scale.}
\end{figure}

\begin{figure}[tbp]
\suppressfloats
\caption{Similar to Fig. 3 for 8.6 keV $^{4}$He atoms impinging on LiF(001)
along the direction \TEXTsymbol{<}100\TEXTsymbol{>}, with the incidence
angle $\protect\theta _{i}=0.71\deg $. Dash-dotted line, first Born
approximation [Eq. (7)]. (a) Full stars, experimental spots, and (b) thick
solid line, experimental intensity, both drawn from Fig. 2 of Ref. [3].}
\end{figure}

\begin{figure}[tbp]
\suppressfloats
\caption{Similar to Fig. 3 for 0.2 keV $^{4}$He atoms impinging on LiF(001)
along the direction \TEXTsymbol{<}110\TEXTsymbol{>}, with the incidence
angle $\protect\theta _{i}=1.5\deg $. Full stars, experimental data from
Fig. 1 of Ref. [2]}
\end{figure}

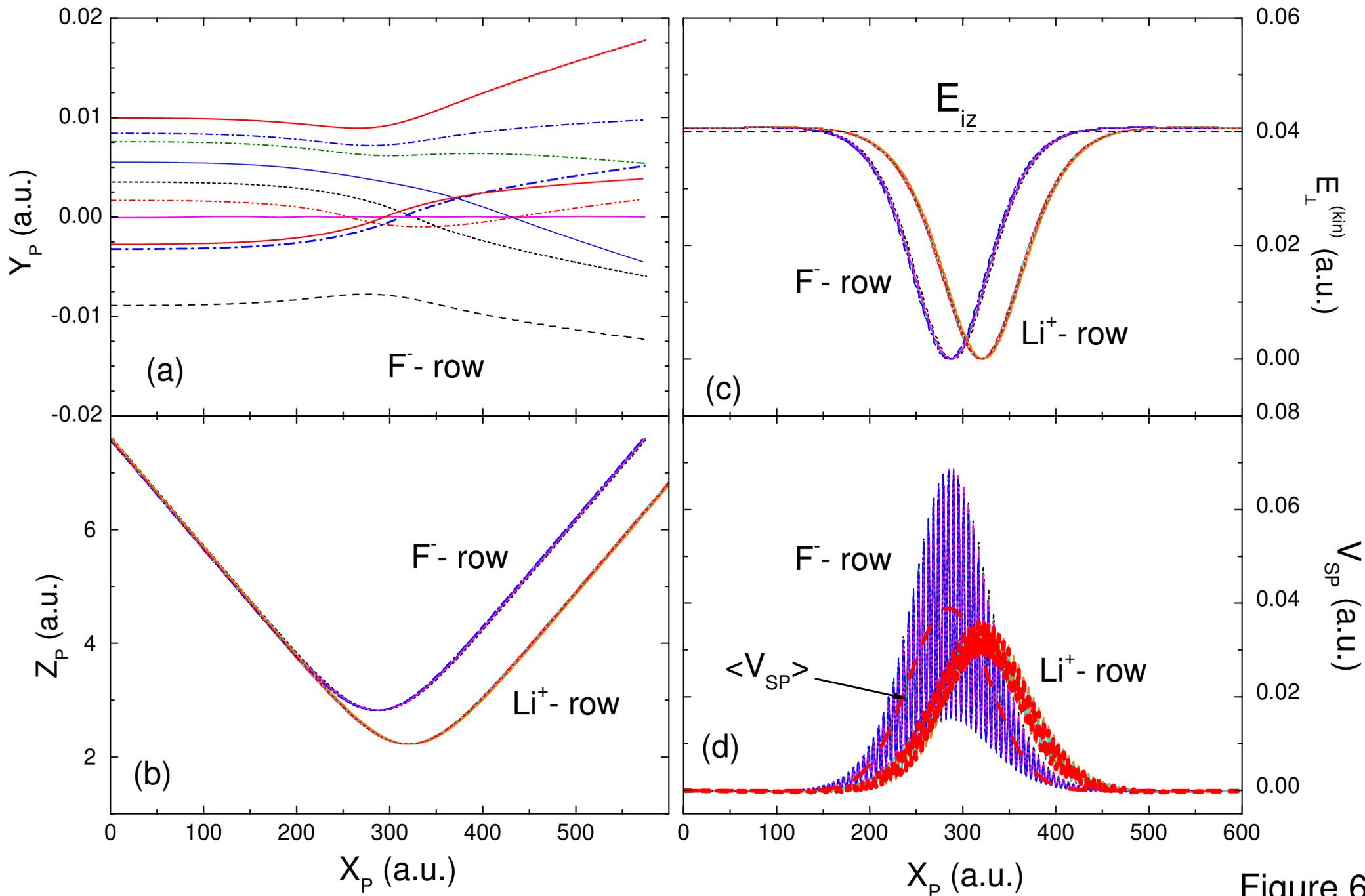
\begin{figure}[tbp]
\suppressfloats
\caption{Classical projectile trajectories ending with a final azimuthal
angle $\protect\varphi _{f}$ near to 0, as a function of the coordinatate $%
X_P$ parallel to the axial channel, for the collision system of Fig. 3. (a)
Coordinate $Y_P$ perpendicular to the scattering plane; (b) coordinate $Z_P$
perpendicular to the surface; (c) transversal kinetic energy along the
trajectory, defined as $E_{\perp}=m_{P}(v_{y}^{2}+v_{z}^{2})/2$; (d)
projectile-surface potential along the trajectory.}
\end{figure}

\end{document}